\documentclass{PoS}
\usepackage{amsmath}
\usepackage{amssymb}
\usepackage{fontenc}
\usepackage{times}
\usepackage{mathptmx}
\usepackage{graphicx}

\title{AdS/QCD at the correlator level}
\ShortTitle{AdS/QCD at the correlator level}

\author{\speaker{Hilmar Forkel} \thanks{%
Supported by the Deutsche Forschungsgemeinschaft.} \\
Institut f\"{u}r Physik, Humboldt-Universit\"{a}t zu Berlin, D-12489 Berlin,
Germany\\
E-mail: forkel@ift.unesp.br}

\abstract{We derive and analyze predictions of both the hard-wall and 
dilaton soft-wall AdS/QCD approximations for the scalar 
glueball correlator and decay constants. We confront the results with 
QCD information from the lattice, the operator product expansion (OPE), 
a hypothetical UV gluon mass associated with the short-distance behavior 
of the heavy-quark potential, and a low-energy theorem based on the 
anomalous dilatational Ward identity. Both duals turn out to encode 
complementary aspects of the above, nonperturbative QCD physics. The OPE 
Wilson coefficients, in particular, are shown to provide a  
challenging testing ground for the impact of the strongly coupled 
holographic UV dynamics on dual gravity predictions.}

\FullConference{8th Conference Quark Confinement and the Hadron Spectrum \\
 September 1-6 2008\\
 Mainz, Germany}

\begin{document}

The first generation of holographic dual candidates for QCD \cite{revs} has
reached a stage of development in which it becomes increasingly important 
to systematically map out their limitations, and thereby to provide a 
quantitative basis for identifying the most promising improvement strategies. 
One way of proceeding in this bottom-up fashion is to survey more detailed 
holographic predictions than the hadron mass spectra. As a step in this 
direction, we have recently derived and analyzed the predicitions of two 
popular AdS/QCD duals, i.e. the hard-wall \cite{pol02} and dilaton soft-wall 
\cite{kar06} backgrounds, for the $0^{++}$ glueball correlation function 
and decay constants \cite{for08}.

Both holographic duals turn out to complement each other in their
representation of specific nonperturbative glueball physics (at momenta
larger than the QCD scale): the soft-wall correlator 
\begin{eqnarray}
\hat{\Pi}^{\rm (sw)}\left( Q^{2}\right) &=&-\frac{2R^{3}}{\kappa ^{2}}
\lambda^{4}\left[
1+\frac{Q^{2}}{4\lambda ^{2}}\left( 1+\frac{Q^{2}}{4\lambda ^{2}}\right)
\psi \left( \frac{Q^{2}}{4\lambda ^{2}}\right) \right]  \notag \\
&&\overset{Q^{2}\gg \lambda ^{2}}{\longrightarrow }-\frac{2}{\pi ^{2}}Q^{4}%
\left[ \ln \frac{Q^{2}}{\mu ^{2}}+\frac{4\lambda ^{2}}{Q^{2}}\ln \frac{Q^{2}%
}{\mu ^{2}}+\frac{2^{2}5}{3}\frac{\lambda ^{4}}{Q^{4}}-\frac{2^{4}}{3}\frac{%
\lambda ^{6}}{Q^{6}}+\frac{2^{5}}{15}\frac{\lambda ^{8}}{Q^{8}}+...\right]
\label{psw}
\end{eqnarray}%
(where $\psi \left( z\right) =\Gamma ^{\prime }\left( z\right) /\Gamma \left(
z\right) $, $\lambda$ is the dilaton mass scale and 
$R^{3}/\kappa ^{2}=2(N_c^2-1)/\pi^2$) 
contains all known types of QCD power corrections, generated
both by vacuum condensates and by a hypothetical UV gluon mass suggested to
encode the short-distance behavior of the static quark-antiquark potential 
\cite{che99}, while sizeable exponential corrections as induced by
small-scale QCD instantons \cite{for01} are reproduced in the hard-wall
correlator%
\begin{eqnarray}
\hat{\Pi}^{\rm (hw)}\left( Q^{2}\right) &=&\frac{R^{3}}{8\kappa ^{2}}Q^{4}
\left[ 2\frac{%
K_{1}\left( Qz_{m}\right) }{I_{1}\left( Qz_{m}\right) }-\ln  \frac{%
Q^{2}}{\mu ^{2}} \right]  \notag \\
&&\overset{Q^{2}\gg z_{m}^{-2}}{\longrightarrow }-\frac{2}{\pi ^{2}}Q^{4}\ln
\frac{Q^{2}}{\mu ^{2}} +\frac{4}{\pi }\left[ 1+\frac{3}{4}%
\frac{1}{Qz_{m}}+O\left( \frac{1}{\left( Qz_{m}\right) ^{2}}\right) \right]
Q^{4}e^{-2Qz_{m}}  \label{phw}
\end{eqnarray}%
(where the IR brane is located at $z_m$). This complementarity generalizes 
to other hadron channels, allows to relate holographic predictions to 
specific aspects of the gauge dynamics and suggests to combine the 
underlying brane- and dilaton-induced IR physics into improved QCD duals.

While the various contributions to the holographic estimates (\ref{psw}) 
and (\ref{phw}) have the expected order of magnitude, the signs of the 
two leading power corrections in Eq. (\ref{psw}) are opposite 
to QCD predictions and violate the factorization approximation to the 
four-gluon condensate. We have argued that this provides specific evidence 
for the short-distance physics in the Wilson coefficients to be inadequately 
reproduced (beyond the leading conformal logarithm) by the strongly-coupled 
UV dynamics of the gravity duals 
\cite{for08}. We have further shown that this problem cannot be mended by 
admixing the UV-subleading solution to the bulk-to-boundary propagator 
(as recently advocated in Ref. \cite{col07}) without loosing consistency 
and predictive power \cite{for208}. It remains to be seen whether 
$\alpha^{\prime }$ corrections, in particular the resummed, local ones 
which are suggested to reproduce the RG flow of the strong coupling 
\cite{gue08}, can generate improved holographic predictions for the power 
corrections.

Since the Wilson coefficients of the $0^{++}$ glueball correlator furthermore
receive unusually small perturbative QCD contributions while those from small 
instantons are enhanced, the hard-wall correlator may produce the better 
overall AdS/QCD prediction. Our holographic estimates for the glueball 
decay constants, which probe aspects of the dual dynamics to which the 
mass spectrum is less sensitive and which are important for experimental 
glueball searches, provide indirect evidence for this expectation. The 
large hard-wall prediction $f_{S}^{\left( \text{hw}\right) }\simeq 
0.8-0.9$ GeV for the ground-state decay constant reflects the strong 
instanton-induced short-distance attraction in the scalar QCD glueball 
correlator, implies an exceptionally small $0^{++}$ glueball size and 
is indeed close to IOPE sum-rule \cite{for01} and lattice \cite{che06} 
results. On the other hand, the absence of instanton contributions in 
the soft wall, which was designed to reproduce confinement-induced 
linear Regge trajectories, agrees with QCD evidence against a direct 
involvement of instantons in the underlying flux-tube formation.

To summarize, our findings demonstrate that the comparison of holographic 
predictions with QCD information at the correlator level provides 
detailed insights into quantitative aspects of the gauge dynamics which 
different gravity duals encode. We have emphasized that information from 
the operator product expansion, which factorizes contributions of short- 
and long-distance physics, allows for a transparent analysis of typical 
shortcomings of holographic models which are 
rooted in their strongly coupled UV sector. These limitations notwithstanding, 
the amount of glueball dynamics which even the simplest holographic duals 
encode is encouraging and indicates that AdS/QCD may indeed 
develop into a systematically improvable approximation to QCD.

We acknowledge financial support from the Funda\c{c}\~{a}o de Amparo a
Pesquisa do Estado de S\~{a}o Paulo (FAPESP) and the Deutsche
Forschungsgemeinschaft (DFG).

\end{document}